\documentclass[10pt, final, twocolumn, twoside, romanappendices]{IEEEtran}

\IEEEoverridecommandlockouts

\usepackage{amsmath,amssymb,amsfonts}
\usepackage{cite}

\usepackage{graphicx}
\usepackage{textcomp}
\usepackage{xcolor}
\usepackage{times}
\usepackage{subfigure}
\usepackage{amssymb,amsmath}
\usepackage{acronym}
\usepackage{balance}

\usepackage{bm}
\usepackage{algorithm}
\usepackage{algorithmic}

\usepackage{lipsum}
\usepackage{stfloats}

\allowdisplaybreaks[4]

\def\BibTeX{{\rm B\kern-.05em{\sc i\kern-.025em b}\kern-.08em
    T\kern-.1667em\lower.7ex\hbox{E}\kern-.125emX}}

\begin{document}

\title{Channel Estimation for RIS Assisted Wireless Communications: Part I - Fundamentals, Solutions, and Future Opportunities \\ {\large {(Invited Paper)}}}

\author{Xiuhong Wei, Decai Shen, and Linglong Dai

\thanks{All authors are with the Beijing National Research Center for Information Science and Technology (BNRist) as well as the Department of Electronic Engineering, Tsinghua University, Beijing 100084, China (e-mails: weixh19@mails.tsinghua.edu.cn, sdc18@mails.tsinghua.edu.cn, daill@tsinghua.edu.cn).}
\thanks{This work was supported in part by the National Key Research and Development Program of China (Grant No. 2020YFB1807201) and in part by the National Natural Science Foundation of China (Grant No. 62031019).}
}

\maketitle

\begin{abstract}
The reconfigurable intelligent surface (RIS) with low hardware cost and energy consumption has been recognized as a potential technique for future 6G communications to enhance coverage and capacity. To achieve this goal, accurate channel state information (CSI) in RIS assisted wireless communication system is essential for the joint beamforming at the base station (BS) and the RIS. However, channel estimation is challenging, since a large number of passive RIS elements cannot transmit, receive, or process signals. In the first part of this invited paper, we provide an overview of the fundamentals, solutions, and future opportunities of channel estimation in the RIS assisted wireless communication system. It is noted that a new channel estimation scheme with low pilot overhead will be provided in the second part of this paper.
\end{abstract}


\vspace{-2mm}
\section{Introduction}\label{S1}
Recently, reconfigurable intelligent surface (RIS) has been proposed to enhance the coverage and capacity of the wireless communication system with low hardware cost and energy consumption~\cite{ZhangEfficiency}. In general, RIS consisting of massive passive low-cost elements can be deployed to establish extra links between the base station (BS) and users. By reconfiguring these RIS elements according to the surrounding environment, RIS can provide high beamforming gain~\cite{RuiZhang19Beamforming}. The reliable beamforming requires accurate channel state information (CSI). Hence, it is essential to develop accurate channel estimation schemes for the RIS assisted wireless communication system~\cite{Qing20Tutorial}.

Although channel estimation has been widely studied in the conventional wireless communication system, there are two main obstacles for conventional schemes to be directly applied in the RIS assisted system~\cite{Power'1}. Firstly, all RIS elements are passive, which cannot transmit, receive, or process any pilot signals to realize channel estimation. Secondly, since an RIS usually consists of hundreds of elements, the dimension of channels to be estimated is much larger than that in conventional systems, which will result in a sharp increase of the pilot overhead for channel estimation. Therefore, channel estimation is a key challenge in the RIS assisted system, which will be investigated in this invited paper composed of two parts.

In the first part, we provide an overview of the fundamentals, solutions, and future opportunities of channel estimation in the RIS assisted wireless communication system. Firstly, the fundamentals of channel estimation are explained in Section II. Then, in Section III, we discuss and compare three types of overhead-reduced channel estimation solutions that exploit the two-timescale channel property, the multi-user correlation, and the channel sparsity, respectively. After that, we point out key challenges and the corresponding future opportunities about channel estimation in the RIS assisted system in Section IV. Finally, some conclusions are drawn in Section V.

{\it Notation}: Lower-case and upper-case boldface letters ${\bf{a}}$ and ${\bf{A}}$ denote a vector and a matrix, respectively; ${{{\bf{A}}^T}}$ and ${{{\bf{A}}^{H}}}$ denote the transpose and conjugate transpose of matrix $\bf{A}$, respectively; ${{\left\|  \bf{a}  \right\|}}$ denotes the ${{l_2}}$-norm of vector ${\bf{a}}$; ${\rm{diag}}\left({\bf{x}}\right)$ denotes the diagonal matrix with the vector $\bf{x}$ on its diagonal.
\vspace{-5mm}

\section{Fundamentals of Channel Estimation in The RIS Assisted System}\label{S2}
In this section, we will first illustrate the system model of an RIS assisted wireless communication system. Then, the channel estimation problem in this system will be presented. Finally, we will introduce the basic channel estimation schemes.
\begin{figure}[tp]
	\begin{center}
		\vspace*{0mm}\includegraphics[width=0.7\linewidth]{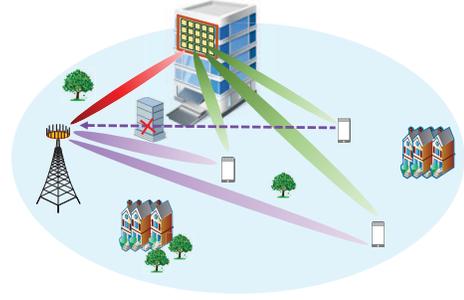}
	\end{center}
	\vspace*{-3mm}\caption{An example of RIS assisted wireless communication system.} \label{FIG1}
	\vspace*{-3mm}
\end{figure}
\subsection{System Model}\label{S2.1}
For the uplink RIS assisted wireless communication system as shown in Fig. \ref{FIG1}, we consider one $M$-antenna BS and one $N$-element RIS to serve $K$ single-antenna users. Let ${\bf{h}}_{d,k}\in{\mathbb{C}}^{M \times 1}$ denote the direct channel between the ${k}$th user and the BS,  ${\bf{G}}\in{\mathbb{C}}^{M \times N}$ be the channel between the RIS and the BS, and ${\bf{h}}_{r,k}\in{\mathbb{C}}^{N \times 1}$ be the channel between the ${k}$th user and the RIS. The received signal ${\bf{y}}\in{\mathbb C}^{M\times1}$ at the BS can be expressed by
\begin{equation}\label{eq17}
{{\bf{y}}} = \sum\limits_{k = 1}^{{K}}\left({\bf{h}}_{d,k}+{\bf{G}}{{\rm{diag}}\left({\bm{\theta}}\right)}{\bf{h}}_{r,k}\right){s}_{k}+{{\bf{w}}},
\end{equation}
where $s_{k}$ is the symbol sent by the $k$th user, $\bm{\theta}=\left[{\theta_{1}},{\theta_{2}},\cdots,{\theta_{N}}\right]^{T}$ is the reflecting vector at the RIS with $\theta_{n}$ representing the reflecting coefficient for the $n$th RIS element, and ${{\bf{w}}}\in\mathbb{C}^{M \times 1}$ is the received noise at the BS. Note that $\theta_{n}$ can be further set as ${\theta_{n}=\beta_ne^{j\phi_n}}$, with $\beta_{n} \in[0,1]$ and $\phi_n \in[0,2 \pi]$ representing the amplitude and the phase for the $n$th RIS element, respectively.

For the RIS assisted system, reliable beamforming requires the accurate CSI consisting of the direct link and the RIS related reflecting link. We consider a time division duplex (TDD) RIS assisted system, where the downlink channel can be obtained based on the estimated uplink channel because of the TDD channel reciprocity.
\vspace{-3mm}

\subsection{Channel Estimation Problem}\label{S2.2}
The channel estimation problem for the direct channel ${\bf{h}}_{d,k}$ can be solved by the conventional schemes in the conventional wireless communication system. Unfortunately, it is difficult to estimate the RIS related channels $\bf{G}$ and ${\bf{h}}_{r,k}$ due to passive RIS elements without signal processing capability.

Let ${\bf{H}}_{k}\triangleq{\bf{G}}{\rm{diag}}\left({\bf{h}}_{r,k}\right)\in{\mathbb{C}}^{M\times N}$ represent the cascaded channel between the $k$th user and the BS via the RIS, and the received signal $\bf{y}$ in~(\ref{eq17}) can be also rewritten as
\begin{equation}\label{eq19}
{{\bf{y}}} = \sum\limits_{k = 1}^{{K}}\left({\bf{h}}_{d,k}+{\bf{H}}_{k}{\bm{\theta}}\right){s}_{k}+{{\bf{w}}}.
\end{equation}
Note that many beamforming algorithms (i.e., how to design the optimal RIS reflecting vector $\bm{\theta}$ in~(\ref{eq19})) aim to optimize the power of the effective reflecting link, i.e., ${\parallel{{\bf{G}}{\rm{diag}}\left({\bm{\theta}}\right){\bf{h}}_{r,k}}\parallel}^2={\parallel{{\bf{G}}{\rm{diag}}\left({\bf{h}}_{r,k}\right){\bm{\theta}}}\parallel}^2={\parallel{{\bf{H}}_{k}{\bm{\theta}}}\parallel}^2$. Therefore, most of existing channel estimation schemes directly estimate the cascaded channel ${\bf{H}}_{k}$ instead of the individual channels $\bf{G}$ and ${\bf{h}}_{r,k}$.

By adopting the orthogonal pilot transmission strategy among users, the uplink channel estimation associated with different users can be independent. Without loss of generality, the subscript $k$ in ${\bf{h}}_{d,k}$, ${\bf{h}}_{r,k}$, and ${\bf{H}}_{k}$ can be omitted to represent the corresponding channels related to any users.
\vspace{-2mm}

\subsection{Basic Channel Estimation Schemes}\label{S2.3}
If all RIS elements are turned off, i.e., the incident electromagnetic wave will be perfectly absorbed by the RIS instead of reflected to the receiver\footnote{Note that ``turn off" is a widely used expression in the literature but inaccurate, since an RIS with all elements turned off is also a scatterer to reflect the incident electromagnetic wave. An implementation method with a special setting of RIS elements proposed in~\cite{Perfect_Absorption} can realize the perfect ``turn off" for the incident electromagnetic wave.}, the RIS assisted communication system can be simplified as the conventional communication system without the RIS. Hence, the direct channel ${\bf{h}}_d$ can be estimated by some classical solutions such as the least square (LS) algorithm.

As mentioned above, the channel estimation for the RIS related channels $\bf{G}$ and ${\bf{h}}_r$ is challenging. A straightforward solution is to estimate the cascaded channel $\bf{H}$ in (\ref{eq19}) based on the ON/OFF protocol proposed in~\cite{Power'1}. The key idea  is to divide the entire cascaded channel estimation process into $N$ stages, where each stage only estimates one column vector of ${\bf{H}}\in{\mathbb{C}}^{M\times N}$ associated with one RIS element. Specifically, the cascaded channel ${\bf{H}}\in{\mathbb{C}}^{M\times N}$ can be represented by $N$ columns as
\begin{equation}\label{eq22}
{\bf{H}}=\left[\mathbf{h}_{1 }, \cdots,\mathbf{h}_{n},\cdots, \mathbf{h}_{N}\right],
\end{equation}
where $\mathbf{h}_{n}\in \mathbb{C}^{M\times 1}$ is the cascaded channel corresponding to the $n$th RIS element. In the $n$th stage, only the $n$th RIS element is turned on, while the remained $N-1$ RIS elements are turned off. Since the direct channel ${\bf{h}}_d$ has been estimated in advance, its impact can be removed from the received pilot signal at the BS. Then, $\mathbf{h}_{n}$ can be estimated based on the LS algorithm. By following this similar procedure, $\mathbf{h}_{1}$, $\cdots$, $\mathbf{h}_{N}$ can be  estimated in turn by sequentially turning on the $1$st, $\cdots$, $N$th RIS element one by one, while  the remained $N-1$ RIS elements are turned off. After $N$ stages, the cascaded channel ${\bf{H}}\in{\mathbb{C}}^{M\times N}$ composed of $N$ columns can be completely estimated. However, since only one RIS element can reflect the pilot signal to the BS based on the ON/OFF protocol, the channel estimation accuracy may be degraded.

In order to improve the channel estimation accuracy,~\cite{Nadeem20DFT} further proposed the discrete Fourier transform (DFT) protocol based channel estimation scheme, where all RIS elements are always turned on. In this scheme, the entire cascaded channel estimation process is still divided into $N$ stages. However, in each stage, the reflecting vector $\bm{\theta}$ at the RIS is specially designed as one column vector of the DFT matrix. After $N$ stages, based on the LS algorithm, the cascaded channel $\bf{H}$ can be directly estimated based on all received pilot signals in $N$ stages at the BS. It is noted that the overall reflecting matrix for $N$ stages forms a DFT matrix of size $N\times N$, which has been proved to be the optimal choice to ensure the channel estimation accuracy~\cite{Nadeem20DFT}.

However, the required pilot overhead in~\cite{Power'1,Nadeem20DFT} is huge. This is mainly caused by the fact that the number of unknown channel coefficients (e.g., $64 \times 256$ with 64 antennas at the BS, 256 elements at the RIS and single antenna at the user) in the RIS assisted communication system is much larger than that of unknown channel coefficients (e.g., $64 \times 1$ with 64 antennas at the BS and a single antenna at the user) in the conventional communication system without the RIS. The huge pilot overhead will significantly decrease the effective capacity improvement. Thus, it is essential to develop overhead-reduced channel estimation schemes for the RIS assisted system. In the next Section \ref{S3}, we will introduce three typical types of overhead-reduced channel estimation solutions.
\vspace{-2mm}

\section{Overhead-Reduced Channel Estimation Solutions}\label{S3}
In this section, we will introduce three typical types of channel estimation solutions to reduce the pilot overhead, which exploit the two-timescale channel property, the multi-user correlation, and the channel sparsity, respectively.
\vspace{-3mm}
\subsection{Two-Timescale Based Channel Estimation}
The first typical solution to reduce the pilot overhead for channel estimation is to exploit the two-timescale channel property in the RIS assisted communication system~\cite{DaiCE1,Anchor-assisted_CE,Matrix-Calibration-Based_CE}.

Specifically, the two-timescale channel property can be explained as follows. On the one hand, since the BS and the RIS are usually placed in fixed positions, the channel $\bf{G}$ between the RIS and the BS usually remains unchanged for a long period of time, which shows the large timescale property. On the other hand, due to the mobility of the user, the channel ${\bf{h}}_r$  between the user and the RIS and the direct channel ${\bf{h}}_d$ between the user and the BS vary in a much smaller timescale than that of the quasi-static channel $\bf{G}$, which show the small timescale property.

As shown in Fig. 2, based on this two-timescale channel property,~\cite{DaiCE1} proposed a two-timescale channel estimation framework, where the two different pilot transmission strategies are respectively designed for estimating the large timescale channel $\bf{G}$ and the small timescale channels ${\bf{h}}_d$ and ${\bf{h}}_r$. Firstly, the high-dimensional channel $\bf{G}$ is estimated once in a large timescale by using the dual-link pilot transmission strategy proposed in~\cite{DaiCE1}. Although the pilot overhead required for estimating $\bf{G}$ is large due to its high dimension, such overhead is negligible from a long-time perspective. Then, based on the widely used uplink pilot transmission strategy, the low-dimensional channels ${\bf{h}}_d$ and ${\bf{h}}_r$ can be estimated before data transmission in a small timescale. Although these channels have to be estimated more frequently, the required pilot overhead is small due to their low dimensions. As a result, the average pilot overhead can be significantly reduced by exploiting the two-timescale channel property.

\begin{figure}[tp]
	\begin{center}
		\vspace*{0mm}\includegraphics[width=0.8\linewidth]{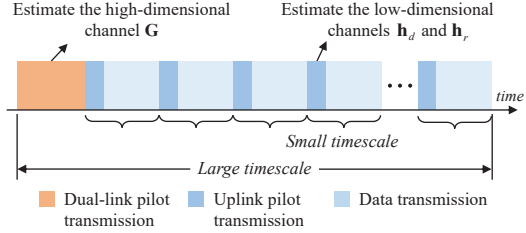}
	\end{center}
	\vspace*{-5mm}\caption{The two-timescale channel estimation framework~\cite{DaiCE1}.} \label{FIG2}
	\vspace*{-3mm}
\end{figure}

The main difficulty of this scheme is how to estimate $\bf{G}$, since all RIS elements are passive without signal processing capability. To achieve this goal,~\cite{DaiCE1} proposed a dual-link pilot transmission strategy as mentioned before. The key idea is that, the BS firstly transmits pilot signals to the RIS via the downlink channel ${\bf{G}}^T$, and then the RIS reflects pilot signals back to the BS via the uplink channel $\bf{G}$. There are $(N+1)$ sub-frames for the dual-link pilot transmission, where each sub-frame consists of $M$ time slots. In the $m_1$th time slot ($m_1=1,2,\cdots,M$) of the $t$th sub-frame ($t=1,2,\cdots,N+1$), the $m_1$th antenna of the BS transmits the pilot signal $s_{m_1,t}$ to the RIS and other $(M - 1)$ antennas of the BS receive the pilot signal reflected by the RIS. The received pilot signal $y_{m_1,m_2,t}$ at the $m_2$th antennas of the BS can be represented as $(m_2=1,2,\cdots,M, m_1\neq m_2)$
\begin{equation}\label{eq23}
\begin{aligned}
y_{m_1,m_2,t}=&\left[{\bf{g}}_{m_2}^T{\rm{diag}}{({\bm{\theta}}_t)}{\bf{g}}_{m_1}+z_{m_1,m_2}\right]s_{m_1,t}+w_{m_1,m_2,t}
\\ = &\left[{\bf{g}}_{m_2}^T{\rm{diag}}{({\bf{g}}_{m_1})}{\bm{\theta}}_t+z_{m_1,m_2}\right]s_{m_1,t}+w_{m_1,m_2,t},
\end{aligned}
\end{equation}
where ${\bf{g}}_{m_2}^T\in{\mathbb{C}}^{1\times N}$ is the $m_2$th row vector of the uplink channel $\bf{G}$, ${\bf{g}}_{m_1}\in{\mathbb{C}}^{N\times 1}$ is the $m_1$th row vector of the downlink channel ${\bf{G}}^T$, ${\bm{\theta}}_t$ is the reflecting vector at the RIS in the $t$th sub-frame, $z_{m_1,m_2}$ is the self-interference after mitigation from the $m_1$th antenna to the $m_2$th antenna of the BS, and $w_{m_1,m_2,t}$ is the received noise at the $m_2$th antenna of the BS. After $N+1$ sub-frames, all received pilot signals $\{y_{m_1,m_2,t}|1\leq m_1,m_2 \leq N, m_1\neq m_2, 1\leq t \leq N+1\}$ can be obtained, which consist of $MN$ unknown variables, i.e., $MN$ elements of  $\bf{G}$. Then, based on all  received pilots, each element of $\bf{G}$ can be alternately  estimated in an iterative manner by utilizing the coordinate descent algorithm~\cite{DaiCE1}.

Some alternative schemes for estimating the channel $\bf{G}$ between the RIS and the BS were proposed~\cite{Anchor-assisted_CE,Matrix-Calibration-Based_CE}. In~\cite{Anchor-assisted_CE}, two users ($U_1$ and $U_2$) are deployed near the RIS to assist its channel estimation. The uplink cascaded channels ${\bf{H}}_{1}$ for user $U_1$, ${\bf{H}}_{2}$ for user $U_2$, and the $U_1$-RIS-$U_2$ cascaded channel between the user $U_1$ and the user $U_2$ via the RIS are estimated based on the pilot symbols transmitted by the two users, respectively. After that, the entries for BS-RIS channel $\bf{G}$ can be calculated based on three estimated cascaded channels mentioned above. By utilizing the long-term channel averaging prior information~\cite{Matrix-Calibration-Based_CE}, the large timescale channel $\bf{G}$ can also be estimated based on the channel matrix calibration.

After acquiring $\bf{G}$, the low-dimensional channels ${\bf{h}}_d$ and ${\bf{h}}_r$ can be estimated based on the conventional uplink pilot transmission strategy. The user transmits the pilot signals to the BS via both the direct channel ${\bf{h}}_d$ and the effective reflecting channel ${\bf{G}}{\bm{\Phi}}{\bf{h}}_r$. Based on the received uplink pilot signals with the known $\bf{G}$ and ${\bm{\Phi}}$, ${\bf{h}}_d$ and ${\bf{h}}_r$ can be directly estimated at the BS by the conventional channel estimation algorithms such as the LS algorithm.

Based on the two-timescale channel property, the large timescale channel $\bf{G}$ and the small timescale channels ${\bf{h}}_d$ and ${\bf{h}}_r$ can be respectively estimated in different timescales, which can indeed significantly reduce the average pilot overhead. However, the channel estimation for $\bf{G}$ is still challenging. In~\cite{DaiCE1}, the BS should work in the  full-duplex mode, where different antennas are required to transmit and receive pilots simultaneously to estimate $\bf{G}$. In~\cite{Anchor-assisted_CE}, the complexity for user scheduling and the overhead for the $U_1$-RIS-$U_2$ cascaded channel feedback are not negligible.

\subsection{Multi-User Correlation Based Channel Estimation}\label{S3.1}
Another solution to reduce the pilot overhead is to directly estimate the corresponding cascaded channels by utilizing the multi-user correlation. Since all users communicate with the BS via the same RIS, the cascaded channels  $\{{{\bf{H}}_k}\}_{k=1}^{K}$ associated with different users have some correlations. Thus, this multi-user correlation can be exploited to reduce the pilot overhead required by the cascaded channel estimation~\cite{Wang20Correlation}.

Specifically, the multi-user correlation can be explained as follows. For convenience, we take the $n$th column $\mathbf{h}_{k,n} \in \mathbb{C}^{M\times 1}$ of the cascaded channel ${\bf{H}}_{k}=\left[\mathbf{h}_{k,1 },\mathbf{h}_{k,2} \cdots, \mathbf{h}_{k,N}\right]\in \mathbb{C}^{M\times N}$ as an example, which can expressed  as
\begin{equation}\label{eq231}
\mathbf{h}_{k,n}=t_{k,n}\mathbf{g}_{n},
\end{equation}
where $t_{k, n}$ denotes the channel between the $k$th user and  the $n$th RIS element, which is also the $n$th element of ${\bf{h}}_r$, ${\bf{g}}_{n} \in \mathbb{C}^{M\times 1}$ denotes the user-independent channel between the $n$th RIS element and the BS, which is also the $n$th column vector of ${\bf{G}}$. Since different users enjoy the same channel ${\bf{G}}$ from the RIS to the BS, $\mathbf{h}_{k,n}$ in (\ref{eq231}) can be rewritten as
\begin{equation}\label{eq24}
\mathbf{h}_{k,n}=\lambda_{k,n}\mathbf{h}_{1,n},
\end{equation}
where
\begin{equation}\label{eq25}
\lambda_{k, n}=\frac{t_{k, n}}{t_{1, n}}.
\end{equation}

The key idea of the multi-user correlation based cascaded channel estimation scheme can be expressed as follows. Firstly, the cascaded channel ${\bf{H}}_{1}=\left[\mathbf{h}_{1,1 }, \cdots,\mathbf{h}_{1,n},\cdots, \mathbf{h}_{1,N}\right]$ for the first user (which is also called as the typical user) can estimated by utilizing the DFT protocol based channel estimation scheme discussed in Subsection \ref{S2.3}. Then, for the $k$th user with $k\geq2$, the column vector $\mathbf{h}_{k,n}$ ($n=1,2,\cdots,N$) of ${\bf{H}}_k$ can be obtained by only estimating the unknown scalar  $\lambda_{k, n}$ in (\ref{eq25}) with only one unknown coefficient instead of $\mathbf{h}_{k,n}$ with $M$ unknown coefficients. Hence, there are only $N$ scalars to be estimated in total  for obtaining the cascaded channel ${\bf{H}}_k$ of size $M\times N$.

By exploiting the multi-user correlation, the pilot overhead can be significantly decreased, since the number of channel coefficients to be estimated becomes much smaller. However, this scheme proposed in~\cite{Wang20Correlation} has assumed that there is no receiving noise at the BS. In the typical scenario of low SNR for channel estimation, the estimation accuracy will be degraded.
\vspace{-3mm}

\subsection{Sparsity Based Channel Estimation}
The overhead-reduced channel estimation solutions in the previous two subsections are mainly realized in the spatial domain. By contrast, in this subsection, we will introduce some overhead-reduced based channel estimation solutions by exploiting the sparsity of channels in the angular domain~\cite{Fang19CS, Liang19CS}.

In the conventional wireless communication system, since there are limited propagation paths, the channel ${\bf{h}}_d$ is sparse in the angular domain. Thus, the channel estimation problem for ${\bf{h}}_d$ can be formulated as a sparse signal recovery problem, which can be solved by compressive sensing (CS) algorithms with reduced pilot overhead. Similarly, the cascaded channel ${\bf{H}}\in\mathbb{C}^{M\times N}$ in RIS assisted systems also shows the sparsity when transformed into the angular domain. Specially, by using the virtual angular-domain representation, the cascaded channel ${\bf{H}}$ can be decomposed as
\begin{equation}\label{eq6}
{{\bf{H}}} = {\bf{U}}_{M}{\tilde{\bf{H}}}{\bf{U}}_{N}^{T},
\end{equation}
where ${\tilde{\bf{H}}}$ denotes the $M\times N$ angular cascaded channel, ${\bf{U}}_{M}$ and ${\bf{U}}_{N}$ are  the ${M \times M}$ and ${N \times N}$ dictionary unitary matrices at the BS and the RIS, respectively. The number of non-zero elements in ${\tilde{\bf{H}}}$ is determined by the product of the number of paths between the RIS and the BS and that between the user and the RIS. Since there are the limited number of scatters around the BS and the RIS, especially in high-frequency communications, ${\tilde{\bf{H}}}$ is usually sparse in nature~\cite{Fang19CS}, as shown in Fig. 3.

\begin{figure}[tp]
	\begin{center}
		\vspace*{0mm}\includegraphics[width=0.8\linewidth]{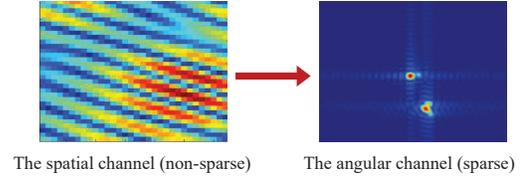}
	\end{center}
	\vspace*{-5mm}\caption{The sparsity of the angular cascaded channel~\cite{Fang19CS}.} \label{FIG3}
	\vspace*{-5mm}
\end{figure}

Based on the sparsity of the angular cascaded channel, the cascaded channel estimation problem can be also formulated as a sparse signal recovery problem~\cite{Fang19CS}. Then, some classical CS algorithms, such as orthogonal matching pursuit (OMP), can be directly used to estimate the angular cascaded channel with reduced pilot overhead. However, these conventional CS algorithms cannot achieve the satisfying estimation accuracy, especially in low SNR ranges. In order to improve the estimation accuracy, a joint sparse matrix recovery based channel estimation scheme was proposed in~\cite{Liang19CS}. In this scheme, all angular channels associated with different users can be projected to the same subspace by considering the fact that different users enjoy the same channel $\bf{G}$ from the RIS to the BS. However, for these sparsity based channel estimation schemes~\cite{Fang19CS,Liang19CS}, the required pilot overhead is still high, since the sparsity of the angular cascaded channel becomes less significant compared with the angular channel in conventional communications.

Moreover,~\cite{RIS_CE_DFT1_ZhangRui} proposed to divide the entire RIS into several sub-surfaces, where all RIS elements on the same sub-surface are considered to have the same channel coefficients. Therefore, the number of the channel coefficients to be estimated can be significantly decreased. By combining the typical overhead-reduced channel estimation schemes mentioned above with this idea of dividing sub-surface, the pilot overhead can be further reduced.

For the sake of simplicity, the above discussions on channel estimation schemes take the narrow band as an example. The similar idea can be extended to the wideband orthogonal frequency division multiplexing (OFDM) channel estimation. Specifically, the channel on each sub-carrier can be estimated separately as in a narrow band system, such as~\cite{RIS_CE_DFT1_ZhangRui}. It is noted that the reflecting vector $\bm{\theta}$ at the RIS are the same for all sub-carriers. Besides, by considering the common sparsity of angular domain channels among different sub-carriers,~\cite{ZhenGao20DL} proposed a joint overhead-reduced channel estimation scheme for all sub-carriers, where a denoising convolution neural network (DnCNN) is used to improve the estimation accuracy.

\section{Challenges and Future Opportunities for Channel Estimation}\label{S4}
In this section, we will point out key challenges for channel estimation in the RIS assisted wireless communication system, based on which the corresponding future research opportunities will be discussed.
\vspace{-2mm}

\subsection{Ultra-Wideband Channel Estimation}\label{S4.2}
In order to achieve ultra-high-speed wireless transmission, the RIS assisted ultra-wideband wireless communication will be an important trend. However, the beam squint effect caused by the ultra-wideband communication brings a serious challenge for the channel estimation, since the single physical angle will be transformed to multiple spatial angles.~\cite{XinyuGao_beamsqilt_TSP} proposed a beam squint pattern matching based wideband channel estimation in the conventional wireless communication system. The similar idea may be applied in the wideband channel estimation for the RIS assisted wireless communication system.
\vspace{-6mm}

\subsection{Spatial Non-Stationarity}
To further exploit the potential beamforming gains and spatial resolutions of the RIS, the number of RIS elements may be hundreds of times larger than that of most scenarios discussed so far, which will result in a large size for the RIS array. With the significant increase of the array size, the RIS related channels will present a new characteristic called as the spatial non-stationarity~\cite{Spatial_Non-stationarity}. This channel characteristic means that the incident direction and power for the electromagnetic wave at the RIS vary with different RIS elements. Under this condition, all existing channel estimation schemes based on the spatial stationarity may not work. The estimation scheme based on the concept of visibility regions~\cite{Spatial_Non-stationarity} may be used to address this challenge.
\vspace{-2mm}

%

\subsection{RIS Assisted Cell-Free Network}
The cell-free network has been recently proposed to address the inter-cell interference of the conventional cellular network. In order to further improve the network capacity with low power consumption, the energy-efficient RISs can be investigated in the cell-free network. However, with the increase of the number of RIS, the number of channels to be estimated increases accordingly in the RIS assisted cell-free network. One possible solution is to exploit the multi-user correlation mentioned in~\cite{Wang20Correlation} to reduce the number of channels to be estimated.
\vspace{-2mm}

\subsection{High Pilot Overhead}\label{S4.4}
Since the RIS consists of a large number of elements, the RIS related channels have many coefficients to be estimated. Although some overhead-reduced channel estimation solutions have been recently proposed, the required pilot overhead is still high due to the passivity of RIS elements. By exploiting more channel characteristics in the RIS, the pilot overhead can be further reduced, which will be discussed in the second part of this invited paper composed of two parts.

\section{Conclusions}\label{S5}
In the first part of this invited paper, we have investigated the channel estimation in the RIS assisted wireless communication system. Due to a large number of passive RIS elements without signal processing capability, the channel estimation in the RIS assisted system is more challenging than that in the conventional system. We first explained the fundamentals of channel estimation. Then, three typical types of overhead-reduced channel estimation solutions were introduced. Finally, we pointed out several challenges and the corresponding future research opportunities for channel estimation. Note that a feasible solution to one of these key challenges, i.e., the high pilot overhead, will be proposed in the second part of this invited paper.


\footnotesize
\balance
\bibliographystyle{IEEEtran}
\bibliography{IEEEabrv,Part_I}

\begin{thebibliography}{10}
\providecommand{\url}[1]{#1}
\csname url@samestyle\endcsname
\providecommand{\newblock}{\relax}
\providecommand{\bibinfo}[2]{#2}
\providecommand{\BIBentrySTDinterwordspacing}{\spaceskip=0pt\relax}
\providecommand{\BIBentryALTinterwordstretchfactor}{4}
\providecommand{\BIBentryALTinterwordspacing}{\spaceskip=\fontdimen2\font plus
\BIBentryALTinterwordstretchfactor\fontdimen3\font minus
  \fontdimen4\font\relax}
\providecommand{\BIBforeignlanguage}[2]{{%
\expandafter\ifx\csname l@#1\endcsname\relax
\typeout{** WARNING: IEEEtran.bst: No hyphenation pattern has been}%
\typeout{** loaded for the language `#1'. Using the pattern for}%
\typeout{** the default language instead.}%
\else
\language=\csname l@#1\endcsname
\fi
#2}}
\providecommand{\BIBdecl}{\relax}
\BIBdecl

\bibitem{ZhangEfficiency}
E.~{Basar}, M.~{Di Renzo}, J.~{De Rosny}, M.~{Debbah}, M.~{Alouini}, and
  R.~{Zhang}, ``Wireless communications through reconfigurable intelligent
  surfaces,'' \emph{IEEE Access}, vol.~7, pp. 116\,753--116\,773, Aug. 2019.

\bibitem{RuiZhang19Beamforming}
Q.~{Wu} and R.~{Zhang}, ``Intelligent reflecting surface enhanced wireless
  network via joint active and passive beamforming,'' \emph{{IEEE} Trans.
  Wireless Commun.}, vol.~18, no.~11, pp. 5394--5409, Nov. 2019.

\bibitem{Qing20Tutorial}
Q.~{Wu}, S.~{Zhang}, B.~{Zheng}, C.~{You}, and R.~{Zhang}, ``Intelligent
  reflecting surface aided wireless communications: A tutorial,'' \emph{arXiv
  preprint arXiv:2007.02759v2}, Jul. 2020.

\bibitem{Power'1}
D.~{Mishra} and H.~{Johansson}, ``Channel estimation and low-complexity
  beamforming design for passive intelligent surface assisted {MISO} wireless
  energy transfer,'' in \emph{Proc. IEEE Int. Conf. Acoust., Speech Signal
  Process. (IEEE ICASSP'19)}, Brighton, UK, May 2019, pp. 4659--4663.

\bibitem{Perfect_Absorption}
M.~F. {Imani}, D.~R. {Smith}, and P.~{Hougne}, ``Perfect absorption in a
  metasurface-programmable complex scattering enclosure,'' \emph{arXiv preprint
  arXiv:2003.01766v2}, Jun. 2020.

\bibitem{Nadeem20DFT}
Q.~{Nadeem}, H.~{Alwazani}, A.~{Kammoun}, A.~{Chaaban}, M.~{Debbah}, and M.~S.
  {Alouini}, ``Intelligent reflecting surface-assisted multi-user {MISO}
  communication: Channel estimation and beamforming design,'' \emph{IEEE Open
  J. Commun. Soc.}, vol.~1, pp. 661--680, May 2020.

\bibitem{DaiCE1}
C.~{Hu} and L.~{Dai}, ``Two-timescale channel estimation for reconfigurable
  intelligent surface aided wireless communications,'' \emph{arXiv preprint
  arXiv:1912.07990}, Dec. 2019.

\bibitem{Anchor-assisted_CE}
X.~{Guan}, Q.~{Wu}, and R.~{Zhang}, ``Anchor-assisted intelligent reflecting
  surface channel estimation for multiuser communications,'' \emph{arXiv
  preprint arXiv:2008.00622}, Aug. 2020.

\bibitem{Matrix-Calibration-Based_CE}
H.~{Liu}, X.~{Yuan}, and Y.~J.~A. {Zhang}, ``Matrix-calibration-based cascaded
  channel estimation for reconfigurable intelligent surface assisted multiuser
  {MIMO},'' \emph{IEEE J. Sel. Areas Commun.}, vol.~38, no.~11, pp. 2621--2636,
  Jul. 2020.

\bibitem{Wang20Correlation}
Z.~{Wang}, L.~{Liu}, and S.~{Cui}, ``Channel estimation for intelligent
  reflecting surface assisted multiuser communications,'' in \emph{Proc. IEEE
  Wireless Commun. and Networking Conf. (IEEE WCNC'20)}, Seoul, Korea, May
  2020, pp. 1--6.

\bibitem{Fang19CS}
P.~{Wang}, J.~{Fang}, H.~{Duan}, and H.~{Li}, ``Compressed channel estimation
  for intelligent reflecting surface-assisted millimeter wave systems,''
  \emph{IEEE Signal Process. Lett.}, vol.~27, pp. 905--909, May 2020.

\bibitem{Liang19CS}
J.~Chen, Y.-C. Liang, H.~V. Cheng, and W.~Yu, ``Channel estimation for
  reconfigurable intelligent surface aided multi-user {MIMO} systems,''
  \emph{arXiv preprint arXiv:1912.03619}, Dec. 2019.

\bibitem{RIS_CE_DFT1_ZhangRui}
B.~{Zheng} and R.~{Zhang}, ``Intelligent reflecting surface-enhanced {OFDM}:
  {Channel} estimation and reflection optimization,'' \emph{IEEE Wireless
  Commun. Lett.}, vol.~9, no.~4, pp. 518--522, Apr. 2020.

\bibitem{ZhenGao20DL}
S.~{Liu}, Z.~{Gao}, J.~{Zhang}, M.~D. {Renzo}, and M.-S. {Alouini}, ``Deep
  denoising neural network assisted compressive channel estimation for {mmWave}
  intelligent reflecting surfaces,'' \emph{IEEE Trans. Veh. Technol.}, vol.~69,
  no.~8, pp. 9223--9228, Jun. 2020.

\bibitem{XinyuGao_beamsqilt_TSP}
X.~{Gao}, L.~{Dai}, S.~{Zhou}, A.~M. {Sayeed}, and L.~{Hanzo}, ``Wideband
  beamspace channel estimation for millimeter-wave {MIMO} systems relying on
  lens antenna arrays,'' \emph{IEEE Trans. Signal Process.}, vol.~67, no.~18,
  pp. 4809--4824, Sep. 2019.

\bibitem{Spatial_Non-stationarity}
A.~{Amiri}, M.~{Angjelichinoski}, E.~{de Carvalho}, and J.~R.~W.~{Heath},
  ``Extremely large aperture massive {MIMO}: Low complexity receiver
  architectures,'' in \emph{Proc. IEEE Globecom Workshops (IEEE GC Wkshps'18)},
  Abu Dhabi, UAE, Dec. 2018, pp. 1--6.

\end{thebibliography}
\end{document}